\documentclass[reprint,amsmath,amssymb,aps,prl,groupedaddress,nofootinbib,twocolumn,superscriptaddress]{revtex4-2}
\usepackage[english]{babel}
\makeatletter
\addto\captionsenglish{\def\l@en{\l@english}}
\makeatother

\usepackage{graphicx}
\usepackage{amsthm,amssymb,amsmath,braket,mathdots}
\usepackage{bm}

\usepackage{graphicx,xcolor} %tikz
\usepackage{tabularx, booktabs} %for tables with width

\input{macros.sty}

\usepackage{pifont}% http://ctan.org/pkg/pifont
\newcommand{\cmark}{\ding{51}}%
\newcommand{\xmark}{\ding{55}}%

\newcommand{\C}{\mathcal{C}_{2}}%some good notation for the C2 rotation in spin space: \C2 = [C_2||1]
\newcommand{\T}{\mathcal{T}}
\renewcommand{\P}{\mathcal{P}}

\newcommand{\sectionPRL}[1]{%
  \refstepcounter{section}% advance section counter
  \phantomsection       % anchor for hyperref
  \addcontentsline{toc}{section}{#1}% adds to TOC
  \par\medskip\noindent%
  \textit{#1.---}\ %
}

\usepackage[caption=false]{subfig}
\usepackage{changes}
\definechangesauthor[name=Valentin, color=blue]{VL} 
\definechangesauthor[name=Johannes, color=red]{JK}

\begin{document}

%\title{Collinear \texorpdfstring{$p$-}{p-}wave magnetism and orbital ferromagnetism from  loop currents}
\title{Collinear \texorpdfstring{$p$-}{p-}wave magnetism and hidden orbital ferrimagnetism}

\author{Valentin Leeb\orcidlink{0000-0002-7099-0682}}
\affiliation{Department of Physics, University of Zürich, Winterthurerstrasse 190, 8057 Zürich, Switzerland}
\author{Johannes Knolle\orcidlink{0000-0002-0956-2419}}
\affiliation{Technical University of Munich, TUM School of Natural Sciences, Physics Department, Garching, Germany}
\affiliation{Munich Center for Quantum Science and Technology (MCQST), Schellingstr. 4, 80799 M{\"u}nchen, Germany}
\affiliation{\small Blackett Laboratory, Imperial College London, London SW7 2AZ, United Kingdom}

\date{\today}

\begin{abstract}
In the absence of spin-orbit coupling, collinear magnets are classified as even-wave magnets, i.e., either ferro-, antiferro-, or altermagnets. It is based on the belief that collinear magnets always feature an inversion-symmetric band structure, which forbids odd-wave magnetism. Here, we show that collinear magnets, which break time reversal symmetry in the non-magnetic sector, can have an inversion symmetry broken band structure and lead to unconventional types of collinear magnets. Hence, collinear odd-wave magnets \emph{do} exist, and we explain that a magnetic field-induced Edelstein effect is their unique signature. We propose minimal models based on the coexistence of AFM order with compensated loop current orders for all types of collinear magnets. Our work provides a new perspective on collinear magnets and the spin-space group classification. 
\end{abstract}
	
\maketitle

\sectionPRL{Introduction}
Collinear magnetic states are conventionally classified in two types: ferromagnets (FMs), which exhibit a finite net magnetization, and antiferromagnets (AFMs), which have compensated magnetic moments. From a symmetry perspective, these two classes differ in how they break time-reversal symmetry (TRS). In FMs, TRS is broken globally and cannot be undone by any real space operation. In AFMs, TRS can be undone by lattice symmetries. In a conventional AFM these lattice symmetries are inversion or translation symmetries, which lead to Kramers degeneracy; the electronic band structure is spin degenerate. Altermagnets (AMs) are discussed as a distinct class of compensated magnets where time reversal is only undone by real-space rotations (or mirror symmetries), resulting in broken Kramers degeneracy \cite{smejkal_conventional_2022,smejkal_emerging_2022}. 
The above classification is only well-defined in the limit where spin and spatial components can be treated separately, i.e., in the absence of spin-orbit coupling, which can be described by the non-relativistic spin-space groups \cite{brinkman_theory_1966, litvin_spin_1974,litvin_spin_1977,corticelli2022spin}. 
%$[\cdot||\cdot]$, in which the first element acts on the spin subspace and the second element acts on the lattice subspace.

\begin{figure*}
    \centering
    \includegraphics[width=\textwidth]{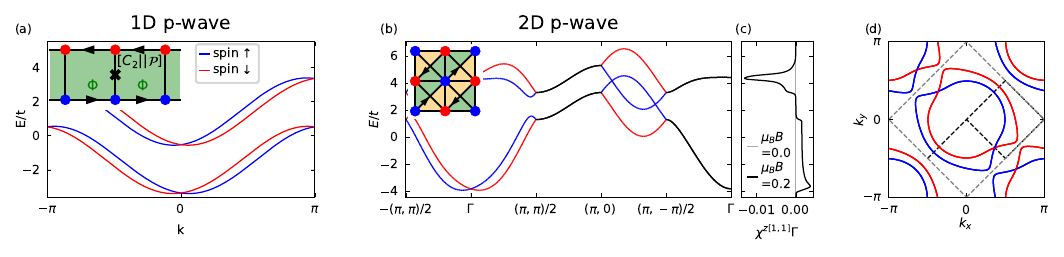}[t]
    \caption{Collinear $p$-wave magnets. (a) Minimal model of an 1D collinear $p$-wave magnet. Inset: An AFM state on a quasi-1D lattice with an orbital magnetic field. Each plaquette carries a flux $\Phi$ which is realized by the complex hoppings, where the current direction is indicated by the arrow. The sign of the magnetization at each site is indicated by the color. The inversion even current pattern reduces the symmetry of the state from $[\mathcal{T}||\mathcal{T}\mathcal{P}]$ to $[\mathcal{T}||\mathcal{P}] = [C_2||\mathcal{P}]$. Panel (a) shows the band structure (color coded by spin) for $\Phi=0.5$ and $m=2t$. (b-d) Stacking the 1D minimal model leads to a minimal model for a 2D collinear AFM. Inset (b): A compensated LCO (arrows indicate current directions, green (orange) plaquettes indicate that the orbital moment is directed out of (into the) plane) coexists with a $(\pi,\pi)$-AFM. (b) The spin-split band structure for $m=z=t$ along the high symmetry path indicated in black in panel (d). (c) The Edelstein susceptibility $\chi^{z[1,1]} = \chi^{zx} + \chi^{zy}$ along the spin-splitting direction is only non-zero in the presence of an in-plane magnetic field. (d) The typical FS of a collinear $p$-wave magnet has a mirror and spin-flip mirror symmetry. }
    \label{fig:1}
\end{figure*}

From a weak coupling perspective in the continuum, Fermi liquid instabilities provide an alternative view on magnetic states~\cite{pomeranchuk_stability_1958,wu_fermi_2007}. A conventional AFM is stabilized by the energy gain from gapping parts of the Fermi surface. This gap can depend on spin and spatial position, and is typically expanded in terms of its angular harmonic $l$. An $l=0$ spin-Pomeranchuk instability then corresponds to a FM as the band structure is spin-split isotropically. Higher even orders of $l$ are classified as AMs as their band structure breaks TRS while featuring no net magnetic moment. Odd-wave magnets, e.g., $p$-wave magnets, have received less attention, but recently several material candidates were suggested \cite{chakraborty_highly_2025,hellenes_unconventional_2024,brekke_minimal_2024} and possibly realized~\cite{yamada2025metallic,song2025electrical,zhou2025anisotropic}. It is widely believed, and a result of the spin-space group classification, that realizing odd-wave magnets requires \emph{non-collinear} magnetic states \cite{smejkal_conventional_2022,brekke_minimal_2024,chakraborty_highly_2025,hellenes_unconventional_2024,yu2025odd},
although recently some counter examples of `odd-parity altermagnets' mostly based on generalizations of the Haldane model were suggested \cite{huang_lightinduced_2025, li_floquet_2025, lin_oddparity_2025,liu_lightinduced_2025,zeng_oddparity_2025,zhu_floquet_2025,zhuang_oddparity_2025,luo_spin_2026}. Here, we demonstrate that this is generally not the case. We explain the symmetry requirements for the emergence of \emph{collinear $p$-wave magnetic states}, give generalizations to orbital ferrimagnets, and provide minimal examples.

Within the spin-space group classification, it has been shown that the non-relativistic band structure of any collinear magnet must be inversion symmetric, which rules out the existence of collinear odd-wave magnets~\cite{smejkal_conventional_2022}. The proof makes the reasonable assumption that the non-magnetic spatial state transforms trivially under time reversal. However, in the context of strongly correlated electron systems, instabilities exist which break TRS spontaneously, without involving the spin degrees of freedom, e.g., loop current orders (LCOs) \cite{affleck_largen_1988,marston_largen_1989,wen_chiral_1989,hsu_two_1991,varma_nonfermiliquid_1997}. Here, we show that the coexistence of a collinear AFM and a non-magnetic state with broken TRS as realized by a LCO realizes inversion broken band structures, e.g., $p$-wave magnets. 

LCOs were suggested in the context of the cuprates as a possible explanation for the pseudo gap \cite{affleck_largen_1988,varma_nonfermiliquid_1997,varma_pseudogap_1999,chakravarty_hidden_2001,varma_theory_2006}. They constitute a type of complex bond order with the expectation value $\Im \langle c_i^\dagger c_j \rangle$ acquiring a finite value resulting in a local current. However, the allowed current patterns are restricted to form loops by the generalized Bloch theorem \cite{bohm_note_1949,watanabe_proof_2019}, which states that an equilibrium ground state must carry zero net current. Even though early polarized-neutron experiments reported evidence for LCOs in the cuprates \cite{fauque_magnetic_2006}, subsequent investigations could not confirm it~\cite{croft_no_2017,fradkin_colloquium_2015}.
LCO are discussed as possible instabilities in other quantum materials aside from the cuprates \cite{bourges_loop_2022}. Famously, the Kagome compound AV$_3$Sb$_5$ shows signatures of TRS-breaking which are typically interpreted as evidence for a LCO \cite{feng_lowenergy_2021,mielke_timereversal_2022,graham_depthdependent_2024}.
From a numerical perspective, LCOs are found in numerous studies, historically inspired by the cuprates \cite{bulut_instability_2015, weber_phase_2014, weber_orbital_2009,kung_numerical_2014} and currently mostly by Kagome systems \cite{tazai_chargeloop_2023,li_intertwined_2024}. 

In this work, we demonstrate that the coexistence of collinear magnets and TRS-breaking of the non-magnetic degrees of freedom, here realized by LCO, leads to inversion symmetry-broken band structures. Crucially, this implies the existence of collinear $p$-wave magnets. Additionally, we discover that the symmetries between spin-up and down bands can be broken entirely, such that a global magnetic moment forms, induced by compensated LCOs. This \textit{hidden orbital ferrimagnet} develops out of a fully compensated AFM.

Our work is organized as follows: First, we refine the symmetry requirements to obtain spin-split band structures in collinear magnets. Next, we explain the symmetry requirements in a minimal 1D model for a collinear $p$-wave magnet. Finally, we generalize this for loop currents on the 2D square lattice and extend this to a minimal model for the cuprates on the Lieb lattice [\ref{A:sec:lieb lattice}].
We note that in our work, we focus on the symmetry aspect of coexisting LCO and AFMs, rather than on microscopic mechanisms for their realization. 

\sectionPRL{Symmetry conditions}
\label{sec:symmetry}
In a collinear magnet without spin-orbit coupling the spin operator commutes with the Hamiltonian. Therefore, we can classify the spin and spatial symmetries separately, which is exactly the idea of the non-relativistic spin-space groups \cite{brinkman_theory_1966, litvin_spin_1974,litvin_spin_1977}. Each element in such a group consists of two parts $[\cdot||\cdot]$, where the first element acts on the spin-subspace and the second element on spatial degrees of freedom. Time reversal $\mathcal{T}$ acts on both elements simultaneously. In this section, we will discuss how to explain the three main features of the electronic band structure of a collinear magnet, i.e., under which conditions the bands are spin-split (i), inversion symmetry broken (ii), and TRS broken (iii). See Tab.~\ref{tab:collinear magnets} for a summary.

First, we review the reasoning of Ref.~\cite{smejkal_conventional_2022} why collinear odd-wave magnets are thought to be impossible. The argument is that for any collinear magnet, time reversal $\mathcal{T}$, followed by a $\C = [C_2||1]$ rotation in spin space, i.e., a $180^\circ$ rotation around an axis perpendicular to the Néel vector, is a symmetry. We denote this as $[\mathcal{T} C_2|| \mathcal{T}] = \C \T$; however, the more common form is $[\mathcal{T} C_2|| 1]$, because $\mathcal{T}$ acts trivially on the real-space coordinates \cite{smejkal_conventional_2022}. Evaluating the effect of $[\mathcal{T} C_2|| 1]$ on the energy bands $\epsilon_s(\vc{k})$, one finds that $[\mathcal{T} C_2|| 1]$ acts effectively like momentum-inversion. Hence, the $[\mathcal{T} C_2|| 1]$ symmetry enforces an inversion symmetric band structure $\epsilon_s(\vc{k}) = \epsilon_s(-\vc{k})$ even in non-centro-symmetric collinear magnets. This also implies the absence of odd-wave spin-splits bands. However, the presence of non-magnetic orders, which transform non-trivially under TRS, causes the argument to break down.

%However, spin and spatial crystal symmetries are insufficient to assess the physical symmetries of a system. First, the non-magnetic state might be coupled to other degrees of freedom like an external field. Secondly, the non-magnetic state itself can also break symmetries spontaneously. Hence, the non-magnetic symmetries of the state must be considered in the above spin-space argument, not just the lattice symmetries. It is now critical to understand that external fields or spontaneous non-magnetic orders can transform non-trivially under time reversal, which directly causes the above argument to break down. The TRS of the non-magnetic part can be broken explicitly, e.g., by a magnetic field, or spontaneously, e.g., by loop currents. 

In collinear magnets the spin degeneracy of the band structure is protected by two independent symmetries because there are two possibilities to invert the spin of an electronic band: A spin rotation $\C=[C_2||1]$ in spin space \emph{and} $[\mathcal{T}|| \mathcal{T} \mathcal{P}] = \T \P$ a combination of time reversal $\T$ and real-space inversion $\P$. Note that spatial and spin inversion can be treated separately, because spin and spatial degrees of freedom are decoupled \cite{smejkal_conventional_2022}. Any of these symmetries may additionally be accompanied by a intra unit cell translation $\vc{t}$, because $[C_2||\vc{t}]$ and $[\mathcal{T}|| \mathcal{T} \mathcal{P} \vc{t}]$ act all in the same way on the electronic bands: They relate bands with opposite spin $\epsilon_\ua(\vc{k}) = \epsilon_\da(\vc{k})$. Hence, both $[C_2|| \vc{t}]$ and $[\mathcal{T}|| \mathcal{T} \mathcal{P} \vc{t}]$ must be broken to obtain spin split bands. Note that this is distinct from the often-encountered understanding that spin inversion/time reversal times any translation or inversion must be broken. The conditions are independent, e.g. an orbital magnetic field, i.e., the part of the magnetic field which couples to the charge degrees of freedom, breaks $\mathcal{T} \mathcal{P}$, because it is a pseudovector, but it does not affect $[C_2|| \vc{t}]$.
%Loop current orders (LCOs), i.e., local spontaneously created magnetic fluxes, can break translation and the $\mathcal{T} \mathcal{P}$ symmetries simultaneously and spontaneously.

The inversion symmetry and the TRS of the band structure of a collinear magnet is also protected by two independent symmetries. For each case the argument is analogous to above.
The explicit spatial inversion $[1||\mathcal{P} \vc{t}]$ but also $[C_2 \mathcal{T}|| \mathcal{T} \vc{t}] = [1|| \mathcal{T} \vc{t}]$ lead to inversion even bands. Hence, non-centro symmetric collinear magnets have an inversion broken band structure if their non-magnetic part transforms non-trivially under $\mathcal{T} \vc{t}$.  Finally, the TRS of the band structure of a collinear magnet is protected by explicit time reversal
$[\mathcal{T}|| \mathcal{T} \vc{t}]$ but also a combination of spin rotation and spatial inversion $[C_2|| \mathcal{P} \vc{t}]$.

It is now insightful to study which magnetic states break the spin-degeneracy, the TRS, and the inversion symmetry of the band structure. The first important thing to realize is that out of the three at least two of these must be broken, because the presence of two symmetries implies the conservation of the third. Therefore, we identify 4 types of unconventional band structures of collinear magnets which the literature typically classifies as different types of magnets: (i) only preserving spin-degeneracy results in an inversion symmetry-broken AFM, (ii) only preserving inversion-symmetry results in an AM (alternatively: even wave magnet excluding $s$-wave), (iii) only preserving TRS results in an odd-wave magnet, and (iv) breaking all 3 symmetries of the band structure results in a \textit{bond ferrimagnet}. Bond ferrimagnets develop a net magnetic moment, due to the inequivalence of the sublattices, even though the spin moments and any orbital moments are individually fully compensated. Tab.~\ref{tab:collinear magnets} summarizes the symmetry requirements. 

The most interesting unconventional collinear magnets are at this point inversion symmetry-broken AFMs and odd-wave magnets, because they are thought to be `forbidden'. Both require the presence of a TRS-broken non-magnetic state. AMs can also emerge from the coexistence of AFMs with time reversal symmetric spontaneous order, like orbital ordering \cite{leeb_spontaneous_2024,daghofer_altermagnetic_2025,meier_antialtermagnetism_2025} or spin loop currents \cite{sato_altermagnetic_2024}. Similarly, bond ferrimagnets may emerge from the coexistence of AFMs with real bond order, which is from a symmetry perspective, identical to the idea of piezomagnetism in AMs \cite{aoyama2024piezomagnetic}, where the application of strain breaks the inversion symmetry of the band structure such that a magnetic moment can form. However, in our examples the magnetic moment is induced by the hidden LCO, which induces compensated orbital moments; hence we will refer to this phenomenon as \emph{hidden orbital ferrimagnetism} --- the generation of an uncompensated magnetic moment due to a hidden orbital order. To summarize this section, the most surprising result of our symmetry analysis is the emergence of collinear odd-wave magnets, because they are generally thought to be forbidden, and will therefore be the focus of our work.

\begin{table}[]
    \centering
    \begin{tabular}{l|c|c|c|l}
         &\multicolumn{3}{c|}{band symmetry}
         \\
         &spin &     inversion&      TRS & example
         \\
         \hline
         conventional AFM
         & \cmark &\cmark&\cmark & Fig.~\ref{fig:LCOs_square}~(a)
         \\
         $\P$-broken AFM
         & \cmark & \xmark & \xmark &  Fig.~\ref{fig:2}~(d), Fig.~\ref{fig:LCOs_lieb}~(a)
         \\
         AM
          & \xmark &\cmark&\xmark & Fig.~\ref{fig:2}~(g), Fig.~\ref{fig:LCOs_lieb}~(g)
         \\
         odd-wave magnet
          & \xmark &\xmark&\cmark & Fig.~\ref{fig:1}, Fig.~\ref{fig:LCOs_lieb}~(c)
         \\
         bond ferrimagnet
          & \xmark &\xmark&\xmark&  Fig.~\ref{fig:2}~(d), Fig.~\ref{fig:LCOs_square}~(c),
          \\
          &&&&Fig.~\ref{fig:LCOs_lieb}~(e)
    \end{tabular}
    \caption{Overview of unconventional collinear magnets. Each of the 3 different fundamental symmetries of the band structure, spin degeneracy, inversion, and TRS, are protected by two different symmetries. They can be broken in 5 ways, as shown here, which leads to 5 types of magnets.}
    \label{tab:collinear magnets}
\end{table}

\sectionPRL{A minimal 1D model}
%We will explain the symmetry requirements in a simple 1D model, where it is trivial to see which symmetries of an AFM need to be broken in order to obtain a $p$-wave spin-split band structure. 
Consider a ladder-like quasi-1D model with a stripe-like AFM state, see Fig.~\ref{fig:1}~(a). There are two sites per unit cell, with opposite spin occupation. The spin degeneracy of the electronic band structure of this state is only protected by the symmetry $[\mathcal{T}||\mathcal{T}\mathcal{P}]$; $[C_2||\vc{t}]$ is not a symmetry, i.e., no translation $\vc{t}$ relates opposite spin sublattices. Hence, any object which transforms odd under $[1||\mathcal{T}\mathcal{P}]$ leads to spin split bands. The easiest example is an orbital magnetic field $\Phi$, i.e., a magnetic field which couples only to the charge of electrons not to the spin (no Zeeman coupling), see Fig.~\ref{fig:1}~(a). The orbital magnetic field falls in this class because it transforms under time reversal but not under inversion because it is a pseudovector, which is reflected in the current pattern in the inset of Fig.~\ref{fig:1}~(a). The band structure is time reversal symmetric, even though $[\T||\T \vc{t}]$ is not a symmetry, because $[C_2||\mathcal{P}]$ is a symmetry of the model. Hence, the band structure 
\begin{equation}
\epsilon_{\pm, s}(k) = -2t \cos k \cos \frac{\Phi}{2} \pm \sqrt{t^2 + (2t \sin k \sin(\Phi/2) +s m)^2}, 
\end{equation}
derived in the end matter and shown in Fig.~\ref{fig:1}~(a), is invariant under $k,s \rightarrow -k, -s$ (here $s=\pm$ is spin up/down), thus it is a collinear $p$-wave magnet.
%
%We expect the 1D example to be, apart from its pedagogical value, of minor experimental relevance, because, first, the spin splitting is controlled by the flux $\Phi$ which is typically small in crystalline systems and comparable to the Zeeman effect. An exception are of course Moire patterns where the unit cell is larger. Secondly, the magnetic field, or in other words an uniformal flux pattern, can only lead in 1D to a $p$-wave magnet. For higher dimensions we therefore focus on LCOs, since they can lead to compensated flux patterns with potentially large fluxes per unit cell. Additionally, LCOs couple intrinsically exclusively to the charge degrees of freedom.

\sectionPRL{Minimal model on the square lattice}
\begin{figure}
    \centering
    \includegraphics[width=\columnwidth]{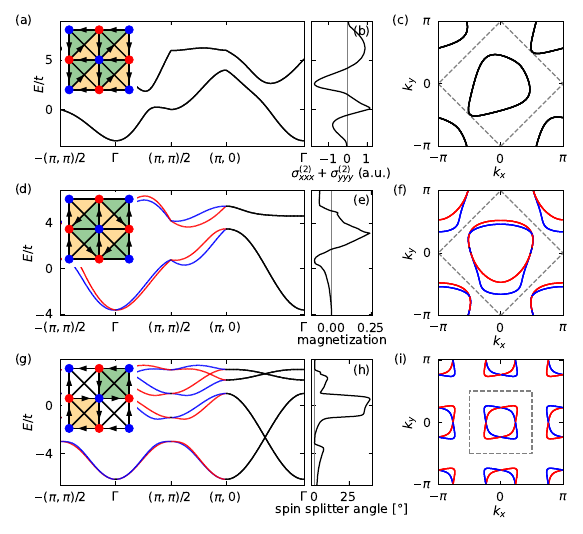}
    \caption{Unconventional LCO-induced magnetic phases in the square lattice. (a,d,g) shows the band structure along a high symmetry line through the Brillouin zone and the inset depicts the LCO and the magnetic state. (c,f,i) shows a representative Fermi surface of this magnetic state. The reduced Brillouin zone (magnetic Brillouin zone) is indicated in grey (dashed). 
    (a-c)
    An inversion-broken AFM. (b) shows the non-reciprocal, second order longitudinal Drude conductivity $\sigma^{(2)}_{xxx}$ (derived in the SM~\cite{supplement}) as characteristic observable, for inversion symmetry breaking.
    (d-f)
    A hidden orbital AFM. (e) shows the magnetization as a characteristic observable, quantifying the spin splitting for various Fermi levels.
    (g-i)
    A LCO-induced AM. (h) shows the spin splitter angle $\alpha$, where $\tan \alpha/2 = (\sigma^\ua_{xy}-\sigma^{\da}_{xy})/(\sigma^\ua_{xx}+\sigma^{\da}_{xx})$ as a characteristic observable which quantifies the spin splitting for various Fermi levels.}
    \label{fig:2}
\end{figure}
Next, we demonstrate the validity of our symmetry analysis by studying different LCOs on the square lattice. We show that all 4 types of unconventional collinear magnetic states can be induced by LCOs. We consider a tight-binding model of spin-full electrons on the square lattice with standard NN hopping $t$ and NNN hopping $t'$ (for concreteness we set $t'=0.5t$) and the dispersion 
\begin{equation}
    \varepsilon(\vc{k}) = -2t \cos k_x -2t \cos k_y - 4t' \cos k_x \cos k_y.
\end{equation}
We add a mean-field AFM collinear order with $\vc{Q} = (\pi,\pi)$ and different LCOs. In a mean-field sense, AFM order acts like a staggered field and LCO as an imaginary hopping $z$. Note that this is \emph{not} spin-orbit coupling, the Hamiltonian remains block-diagonal in spin and reads
\begin{align}
    H &= \sum_{\vc{k},s} \begin{pmatrix} c_{\vc{k},s}\\ c_{\vc{k}+\vc{Q},s} \end{pmatrix}^\dag
    h(\vc{k})
    \begin{pmatrix} c_{\vc{k},s}\\ c_{\vc{k}+\vc{Q},s} \end{pmatrix}
    \\
    h(\vc{k}) &=
    \begin{pmatrix}
    \varepsilon(\vc{k}) & s m \\
    s m & \varepsilon(\vc{k}+\vc{Q})
    \end{pmatrix}
    + \vc{\sigma} \cdot \vc{d}(\vc{k})
    \label{eq:H_square}
\end{align}
where $\vc{\sigma} = (\sigma_0,\sigma_x,\sigma_y,\sigma_z)$ are the Pauli matrices and $\vc{d}(\vc{k})$ includes the contributions of the loop currents.

The bands and eigenstates of the Hamiltonian \eqref{eq:H_square} can be calculated analytically
\begin{equation}
    \epsilon_{\pm,s}(\vc{k}) = 
    \varepsilon_+ \pm \sqrt{(d_x+ (-1)^s m)^2 + d_y^2 + (\varepsilon_- +d_z)^2}
    \label{eq:bands_square}
\end{equation}
where $\varepsilon_\pm(\vc{k}) = (\varepsilon(\vc{k}) \pm \varepsilon(\vc{k}+\vc{Q}))/2$ and we dropped $\vc{k}$ dependencies in \eqref{eq:bands_square} for convenience. The analytic expression for the electronic bands shows that $d_x \neq 0$ is the requirement to obtain spin-split bands. 
In the following, we show four different examples: a collinear, LCO-induced $p$-wave magnet [Fig.~\ref{fig:1}~(b)]; an inversion symmetry-broken AFM [Fig.~\ref{fig:2}~(a)]; a hidden orbital ferrimagnet [Fig.~\ref{fig:2}~(b)]; and an AM [Fig.~\ref{fig:2}~(c)]. We provide a simple analytic description of the first three which are induced by $\vc{Q}=(0,0)$ and $\vc{Q}=(\pi,\pi)$ LCOs. 

The $p$-wave magnet [Fig.~\ref{fig:1}~(b)] is induced by the staggered version of the above LCO. Note,  it can also be considered the compensated 2D analog of our minimal 1D example, see inset. The $p$-wave like $(\pi,\pi)$-LCO with a mirror symmetry in $[1,1]$ direction is described by 
\begin{align}
    \vc{d}_{p^2_{[1,1]}} (\vc{k}) &= 2 z_{p^2_{[1,1]}} \sin( k_x + k_y ) \hat{\vc{e}}_x.
\end{align}
Because the AFM and the LCO have both non-zero elements in the $x$-component of the Bloch vector, the bands are spin split but invariant under $s,\vc{k} \rightarrow -s, -\vc{k}$, resulting in $p$-wave spin-splitting along the $[1,1]$ direction. The spin splitting can be quantified by the magnetic Edelstein effect, see Fig.~\ref{fig:1}~(c), as we discuss in the next section.

The inversion symmetry-broken AFM [Fig.~\ref{fig:2}~(a)] is induced by a $p$-wave like $(0,0)$-LCO with a mirror symmetry in $[1,1]$ direction, see the inset of Fig.~\ref{fig:2}~(a).
Hence, it is described by 
\begin{align}
    \vc{d}_{p^1_{[1,1]}} (\vc{k}) &= 2 z_{p^1_{[1,1]}}  \left[\sin k_x + \sin k_y - \sin (k_x+k_y) \right] \hat{\vc{e}}_0.
\end{align}
Therefore, the bands $\epsilon_{\pm,s}(\vc{k})$ are only invariant under $s \rightarrow -s $ and $(k_x,k_y) \rightarrow (k_y,k_x)$,  which is an inversion symmetry-broken spin degenerate band structure.

The hidden orbital ferrimagnet [Fig.~\ref{fig:2}~(d)] is induced by a $(\pi,\pi)$-LCO with a mirror axis in $[0,1]$ direction, see the inset of Fig.~\ref{fig:2}~(b).  It is described by
\begin{equation}
    \vc{d}_{p^2_{[0,1]}} (\vc{k}) = z_{p^2_{[0,1]}}
    \begin{pmatrix}
        \sin(k_x+k_y) - \sin(k_x-k_y)
        \\
        \sin(k_x-k_y) - \sin(k_x+k_y)
        \\
        0
        \\
        -2 \sin k_y
    \end{pmatrix},
\end{equation}
which results in bands which are only mirror invariant $(k_x,k_y) \rightarrow (-k_x,k_y)$. Because there is no symmetry which relates bands with opposite spin (or in real space the spin sublattices are not symmetry equivalent), a net magnetic moment can form like in a ferrimagnet. Nevertheless, the spin degeneracy at certain high symmetry points in the Brillouin zone, like the $\Gamma$-point or the Brillouin zone boundary is symmetry protected. Therefore, the fundamental difference to a conventional ferromagnet is that bands are not spin-split at every energy, instead the magnetization builds up due to spin-dependent effective masses, see Fig.~\ref{fig:2}~(e).

Finally, the co-existence of LCO with AFM can also induce an AM [Fig.~\ref{fig:2}~(g)]. We choose a $(\pi,0)+(0,\pi)$-LCO, i.e., it has a $2 \times 2$ unit cell, see inset. Hence, it can not be described by \eqref{eq:H_square} and its Bloch Hamiltonian is provided in the SM~\cite{supplement}. We quantify the spin splitting of an altermagnetic Fermi surface by the spin splitter effect \cite{gonzalez-hernandez_efficient_2021}, see Fig.~\ref{fig:2}~(h).

In the end matter, we provide additional minimal models on the square lattice. Moreover, we show that the Lieb lattice, related to the Emery model of the cuprates, provides an ideal basis for LCO-induced unconventional magnets, because the more complex unit cell enriches the possibilities of LCO symmetries.

\sectionPRL{Magnetic Edelstein effect}
The Edelstein effect refers to a non-equilibrium spin accumulation $\delta S^i$ generated by an applied electric field $E^j$ \cite{edelstein_spin_1990}. On a phenomenological level, such a linear coupling between current and spin --- corresponding to a non-vanishing magneto-electric tensor --- has been known for decades \cite{dzyaloshinskii_magnetoelectrical_1960,gorkov_kinetic_1987}. Within linear-response theory, treating the electric field as a perturbation and the spin density as the resulting observable, the response is given by the Kubo formula
\begin{align}
    \delta S^i =& \frac{\Re}{2\pi}  \sum_{\vc{k},n,m}   \left[G_n(\vc{k})G_m^*(\vc{k}) - G_n(\vc{k}) G_m(\vc{k}) \right]
    \label{eq:Edelstein suscetibility}
    \\
    &\times \langle u_n(\vc{k}) | \sigma^i|u_m(\vc{k})\rangle  \langle u_m(\vc{k}) | e E^j\partial_j h(\vc{k}) |u_n(\vc{k})\rangle
    \nonumber
\end{align}
where $G_n^{-1} = E_n(\vc{k}) + \ii \Gamma$ is the retarded Green's function of band $n$, broadened by a finite inverse lifetime $\Gamma$ \cite{chakraborty_highly_2025}.
Here, we evaluate the Edelstein susceptibility $\chi^{ij} \Gamma =\delta S^i \Gamma/V  E^j$, which typically scales with the inverse lifetime $\Gamma$. 

For any spin-commuting Hamiltonian $[\sigma^i,H] = 0$ the Edelstein susceptibility vanishes; we prove this in the end matter~\ref{A:sec:Edelstein effect 0}. Consequently, collinear $p$-wave magnets do not exhibit an Edelstein effect. The underlying reason is that for spin-commuting Hamiltonians, the spin matrix elements are diagonal, $\langle u_n(\vc{k}) | \sigma^i|u_m(\vc{k})\rangle \propto \delta_{n,m}$, such that \eqref{eq:Edelstein suscetibility} reduces to a simple Fermi surface integration of the Fermi velocity, which averages to zero. Already the contribution of a single Fermi surface sheet vanishes.

A finite Edelstein response, however, can be induced by applying a magnetic (Zeeman) field perpendicular to the Néel vector; we refer to this as the magnetic Edelstein effect. This response is not generic for all collinear $p$-wave magnets, because it depends sensitively on the Fermi surface: it requires an unprotected crossing of spin-up and spin-down bands at the Fermi level. For example in Fig.~\ref{fig:1}~(c) around $\mu=2$, the magnetic Edelstein susceptibility vanishes, because the only band crossings lie at the Brillouin zone boundary, where the spin degeneracy is protected by $[C_2||M_{[1,-1]}]$.

%When present, the magnetic Edelstein susceptibility is a smoking gun signature of a collinear $p$-wave magnet. 
The vanishing Edelstein susceptibility at zero magnetic field, which is switched on by finite transverse fields, sets collinear $p$-wave magnets apart from conventional $p$-wave magnets. The tensor structure of $\chi^{ij}$ follows the symmetry of the $p$-wave magnet, i.e., $|\chi^{ij}|$ is maximal when the electric field $\vc{E}$ points along the direction of spin splitting, and $\chi^{ij}=0$ when $\vc{E}$ is aligned with the nodal directions. As expected from symmetry, the in-plane orientation of the magnetic field (within the plane perpendicular to the Néel vector) does not affect the response.
We compute the magnetic Edelstein susceptibility for the minimal square lattice $p$-wave magnet in Fig.~\ref{fig:1}~(c). 

\sectionPRL{Conclusion}
We analyze the symmetry conditions to obtain spin-split band structures in collinear magnets without spin-orbit coupling and found 4 types of unconventional collinear magnets. Most importantly, we showed that the presence of TRS breaking LCO leads to collinear odd-wave magnets and inversion-symmetry broken AFMs. The former  are characterized by TRS and spin splitting of the bands, whereas the bands of the latter have no TRS but are spin degenerate. The band structure can even break TRS, inversion, and spin degeneracy simultaneously which leads to a hidden orbital ferrimagnet, in which a net moment is induced by the compensated loop current. We provide minimal models for all types of magnets on the square lattice (and the Lieb lattice in the SM). Finally, we explained that our central finding, collinear $p$-wave magnets, can be uniquely identified by the magnetic Edelstein effect, which sets them apart from non-collinear $p$-wave magnets. 

The collinear inversion symmetry-broken AFM (odd-wave magnet) fulfills all symmetry conditions to experience a quantum anomalous Hall effect (quantum anomalous spin Hall effect), even in the absence of spin orbit coupling. This is highly uncommon, because these effects are typically associated \emph{with} spin-orbit coupling. However, we observe that all of our minimal models have a vanishing Berry curvature. We dedicate this to the simplicity of our minimal models which are focused on exact solubility. In general, we expect collinear inversion symmetry-broken AFM and hidden orbital ferrimagnets to experience a quantum anomalous Hall effect, and collinear odd-wave magnets to show a quantum anomalous spin Hall effect. Understanding the microscopic mechanism for their appearance will be an important direction for future research. 

Finally, we would like to highlight that the emergence of unconventional collinear magnets arises generically from the coexistence of AFM  with other non-magnetic but TRS-breaking orders. It is \emph{not} constrained to the LCO. Other mechanisms like adatom engineering~\cite{pupim_adatom_2025}, the coexistence with chiral superconductors \cite{khim_coexistence_2025,kuboki_coexistence_2010}, or fluctuating loop currents \cite{palle_superconductivity_2024}, as well as strong magnetic fields in quasi-1D Moire systems may provide alternative routes towards experimental realization of collinear $p$-wave magnets. Hence, the microscopic realization of such states remains an outstanding problem.

%Even though there are no direct experimental realizations in sight, the ideas of our work fill a major knowledge gap in field of unconventional magnets.

% \section*{Data and code availability}
% Code and data related to this paper are available on Zenodo CODE from the authors
% upon reasonable request.

\begin{acknowledgments}
V.L. thanks B.E.~Lüscher and S.~Castro Holbæk for helpful discussions. We thank R. Fernandes, A. Mook, and L. Smejkal for insightful discussions. 
J.K. acknowledges support from the Deutsche Forschungsgemeinschaft (DFG, German Research Foundation) under Germany’s Excellence Strategy (EXC–2111–390814868), and DFG Grants
No. KN1254/1-2, KN1254/2-1 TRR 360 – 492547816 [14] and SFB 1143 (project-id 247310070), as well as the Munich Quantum Valley, which is supported by the Bavarian
state government with funds from the Hightech Agenda Bayern Plus. J.K. further acknowledges support from the Imperial-TUM flagship partnership.
\end{acknowledgments}

%\clearpage
\bibliography{zotero,extra_ref} %bib

@article{li_exploring_2025,
	title = {Exploring \$d\$-wave magnetism in cuprates from oxygen moments},
	volume = {112},
	url = {https://link.aps.org/doi/10.1103/vx12-r2k1},
	doi = {10.1103/vx12-r2k1},
	number = {12},
	urldate = {2025-11-26},
	journal = {Physical Review B},
	author = {Li, Ying and Leeb, Valentin and Wohlfeld, Krzysztof and Valentí, Roser and Knolle, Johannes},
	month = sep,
	year = {2025},
	note = {Publisher: American Physical Society},
	pages = {125139},
}

@article{emery_theory_1987,
	title = {Theory of high-\$\{{\textbackslash}mathrm\{{T}\}\}\_\{{\textbackslash}mathrm\{c\}\}\$ superconductivity in oxides},
	volume = {58},
	url = {https://link.aps.org/doi/10.1103/PhysRevLett.58.2794},
	doi = {10.1103/PhysRevLett.58.2794},
	number = {26},
	urldate = {2025-11-26},
	journal = {Physical Review Letters},
	author = {Emery, V. J.},
	month = jun,
	year = {1987},
	note = {Publisher: American Physical Society},
	pages = {2794--2797},
}

@article{palle_superconductivity_2024,
	title = {Superconductivity due to fluctuating loop currents},
	volume = {10},
	url = {https://www.science.org/doi/10.1126/sciadv.adn3662},
	doi = {10.1126/sciadv.adn3662},
	number = {24},
	urldate = {2025-11-26},
	journal = {Science Advances},
	author = {Palle, Grgur and Ojajärvi, Risto and Fernandes, Rafael M. and Schmalian, Jörg},
	month = jun,
	year = {2024},
	note = {Publisher: American Association for the Advancement of Science},
	pages = {eadn3662},
}

@article{kuboki_coexistence_2010,
	series = {Proceedings of the 9th {International} {Conference} on {Materials} and {Mechanisms} of {Superconductivity}},
	title = {Coexistence of antiferromagnetism and \textit{d}-wave superconductivity in extended \textit{t}–\textit{{J}} model},
	volume = {470},
	issn = {0921-4534},
	url = {https://www.sciencedirect.com/science/article/pii/S0921453409009502},
	doi = {10.1016/j.physc.2009.12.053},
	urldate = {2025-11-26},
	journal = {Physica C: Superconductivity and its Applications},
	author = {Kuboki, Kazuhiro and Yoneya, Masanao and Yamase, Hiroyuki},
	month = dec,
	year = {2010},
	pages = {S163--S164},
}

@article{khim_coexistence_2025,
	title = {Coexistence of local magnetism and superconductivity in the heavy-fermion compound \$\{{\textbackslash}mathrm\{{CeRh}\}\}\_\{2\}\{{\textbackslash}mathrm\{{As}\}\}\_\{2\}\$ revealed by \${\textbackslash}ensuremath\{{\textbackslash}mu\}{\textbackslash}mathrm\{{SR}\}\$ studies},
	volume = {111},
	url = {https://link.aps.org/doi/10.1103/PhysRevB.111.115134},
	doi = {10.1103/PhysRevB.111.115134},
	number = {11},
	urldate = {2025-11-26},
	journal = {Physical Review B},
	author = {Khim, Seunghyun and Stockert, Oliver and Brando, Manuel and Geibel, Christoph and Baines, Chirstopher and Hicken, Thomas J. and Luetkens, Hubertus and Das, Debarchan and Shiroka, Toni and Guguchia, Zurab and Scheuermann, Robert},
	month = mar,
	year = {2025},
	note = {Publisher: American Physical Society},
	pages = {115134},
}

@article{pupim_adatom_2025,
	title = {Adatom {Engineering} {Magnetic} {Order} in {Superconductors}: {Applications} to {Altermagnetic} {Superconductivity}},
	volume = {134},
	shorttitle = {Adatom {Engineering} {Magnetic} {Order} in {Superconductors}},
	url = {https://link.aps.org/doi/10.1103/PhysRevLett.134.146001},
	doi = {10.1103/PhysRevLett.134.146001},
	number = {14},
	urldate = {2025-11-26},
	journal = {Physical Review Letters},
	author = {Pupim, Lucas V. and Scheurer, Mathias S.},
	month = apr,
	year = {2025},
	note = {Publisher: American Physical Society},
	pages = {146001},
}

@article{gorkov_kinetic_1987,
	title = {Kinetic {Effects} in {Antiferromagnetic} {Conductors} with a {Spin}-{Density} {Wave}},
	volume = {93},
	issn = {0044-4510},
	language = {Russian},
	number = {6},
	journal = {Zhurnal Eksperimentalnoi I Teoreticheskoi Fiziki},
	author = {Gorkov, Lp and Sokol, Av},
	month = dec,
	year = {1987},
	note = {Num Pages: 13
Place: Moscow
Publisher: Mezhdunarodnaya Kniga
Web of Science ID: WOS:A1987L487500027},
	pages = {2219--2231},
}

@article{edelstein_spin_1990,
	title = {Spin polarization of conduction electrons induced by electric current in two-dimensional asymmetric electron systems},
	volume = {73},
	issn = {0038-1098},
	url = {https://www.sciencedirect.com/science/article/pii/003810989090963C},
	doi = {10.1016/0038-1098(90)90963-C},
	number = {3},
	urldate = {2025-11-26},
	journal = {Solid State Communications},
	author = {Edelstein, V. M.},
	month = jan,
	year = {1990},
	pages = {233--235},
}

@article{wen_chiral_1989,
	title = {Chiral spin states and superconductivity},
	volume = {39},
	url = {https://link.aps.org/doi/10.1103/PhysRevB.39.11413},
	doi = {10.1103/PhysRevB.39.11413},
	number = {16},
	urldate = {2025-11-25},
	journal = {Physical Review B},
	author = {Wen, X. G. and Wilczek, Frank and Zee, A.},
	month = jun,
	year = {1989},
	note = {Publisher: American Physical Society},
	pages = {11413--11423},
}

@article{hsu_two_1991,
	title = {Two observable features of the staggered-flux phase at nonzero doping},
	volume = {43},
	url = {https://link.aps.org/doi/10.1103/PhysRevB.43.2866},
	doi = {10.1103/PhysRevB.43.2866},
	number = {4},
	urldate = {2025-11-25},
	journal = {Physical Review B},
	author = {Hsu, T. C. and Marston, J. B. and Affleck, I.},
	month = feb,
	year = {1991},
	note = {Publisher: American Physical Society},
	pages = {2866--2877},
}

@article{brinkman_theory_1966,
	title = {Theory of spin-space groups},
	copyright = {Scanned images copyright © 2017, Royal Society},
	url = {https://royalsocietypublishing.org/doi/10.1098/rspa.1966.0211},
	doi = {10.1098/rspa.1966.0211},
	language = {EN},
	urldate = {2025-11-18},
	journal = {Proceedings of the Royal Society of London. Series A. Mathematical and Physical Sciences},
	author = {Brinkman, W. F. and Elliott, Roger James},
	month = oct,
	year = {1966},
	note = {Publisher: The Royal SocietyLondon},
}

@article{litvin_spin_1977,
	title = {Spin point groups},
	volume = {33},
	issn = {0567-7394},
	url = {https://journals.iucr.org/a/issues/1977/02/00/a14103/},
	doi = {10.1107/S0567739477000709},
	language = {en},
	number = {2},
	urldate = {2025-11-18},
	journal = {Acta Crystallographica Section A: Crystal Physics, Diffraction, Theoretical and General Crystallography},
	author = {Litvin, D. B.},
	month = mar,
	year = {1977},
	note = {Publisher: International Union of Crystallography},
	pages = {279--287},
}

@article{litvin_spin_1974,
	title = {Spin groups},
	volume = {76},
	issn = {0031-8914},
	url = {https://www.sciencedirect.com/science/article/pii/0031891474901578},
	doi = {10.1016/0031-8914(74)90157-8},
	number = {3},
	urldate = {2025-11-18},
	journal = {Physica},
	author = {Litvin, D. B. and Opechowski, W.},
	month = sep,
	year = {1974},
	pages = {538--554},
}

@article{li_intertwined_2024,
	title = {Intertwined {Van} {Hove} {Singularities} as a {Mechanism} for {Loop} {Current} {Order} in {Kagome} {Metals}},
	volume = {132},
	url = {https://link.aps.org/doi/10.1103/PhysRevLett.132.146501},
	doi = {10.1103/PhysRevLett.132.146501},
	number = {14},
	urldate = {2025-11-18},
	journal = {Physical Review Letters},
	author = {Li, Heqiu and Kim, Yong Baek and Kee, Hae-Young},
	month = apr,
	year = {2024},
	note = {Publisher: American Physical Society},
	pages = {146501},
}

@article{fradkin_colloquium_2015,
	title = {Colloquium: {Theory} of intertwined orders in high temperature superconductors},
	volume = {87},
	shorttitle = {Colloquium},
	url = {https://link.aps.org/doi/10.1103/RevModPhys.87.457},
	doi = {10.1103/RevModPhys.87.457},
	number = {2},
	urldate = {2025-11-18},
	journal = {Reviews of Modern Physics},
	author = {Fradkin, Eduardo and Kivelson, Steven A. and Tranquada, John M.},
	month = may,
	year = {2015},
	note = {Publisher: American Physical Society},
	pages = {457--482},
}

@article{croft_no_2017,
	title = {No evidence for orbital loop currents in charge-ordered \$\{{\textbackslash}mathrm\{{YBa}\}\}\_\{2\}\{{\textbackslash}mathrm\{{Cu}\}\}\_\{3\}\{{\textbackslash}mathrm\{{O}\}\}\_\{6+x\}\$ from polarized neutron diffraction},
	volume = {96},
	url = {https://link.aps.org/doi/10.1103/PhysRevB.96.214504},
	doi = {10.1103/PhysRevB.96.214504},
	number = {21},
	urldate = {2025-11-18},
	journal = {Physical Review B},
	author = {Croft, T. P. and Blackburn, E. and Kulda, J. and Liang, Ruixing and Bonn, D. A. and Hardy, W. N. and Hayden, S. M.},
	month = dec,
	year = {2017},
	note = {Publisher: American Physical Society},
	pages = {214504},
}

@article{watanabe_proof_2019,
	title = {A {Proof} of the {Bloch} {Theorem} for {Lattice} {Models}},
	volume = {177},
	issn = {1572-9613},
	url = {https://doi.org/10.1007/s10955-019-02386-1},
	doi = {10.1007/s10955-019-02386-1},
	language = {en},
	number = {4},
	urldate = {2025-11-18},
	journal = {Journal of Statistical Physics},
	author = {Watanabe, Haruki},
	month = nov,
	year = {2019},
	pages = {717--726},
}

@article{bohm_note_1949,
	title = {Note on a {Theorem} of {Bloch} {Concerning} {Possible} {Causes} of {Superconductivity}},
	volume = {75},
	url = {https://link.aps.org/doi/10.1103/PhysRev.75.502},
	doi = {10.1103/PhysRev.75.502},
	number = {3},
	urldate = {2025-11-18},
	journal = {Physical Review},
	author = {Bohm, D.},
	month = feb,
	year = {1949},
	note = {Publisher: American Physical Society},
	pages = {502--504},
}

@misc{hellenes_unconventional_2024,
	title = {Unconventional p-wave magnets},
	url = {http://arxiv.org/abs/2309.01607},
	doi = {10.48550/arXiv.2309.01607},
	urldate = {2025-11-18},
	publisher = {arXiv},
	author = {Hellenes, Anna Birk and Jungwirth, Tomáš and Sinova, Jairo and Šmejkal, Libor},
	month = mar,
	year = {2024},
	note = {arXiv:2309.01607 [cond-mat]
version: 2},
}

@article{garate_nonadiabatic_2009,
	title = {Nonadiabatic spin-transfer torque in real materials},
	volume = {79},
	copyright = {http://link.aps.org/licenses/aps-default-license},
	issn = {1098-0121, 1550-235X},
	url = {https://link.aps.org/doi/10.1103/PhysRevB.79.104416},
	doi = {10.1103/PhysRevB.79.104416},
	language = {en},
	number = {10},
	urldate = {2025-11-10},
	journal = {Physical Review B},
	author = {Garate, Ion and Gilmore, K. and Stiles, M. D. and MacDonald, A. H.},
	month = mar,
	year = {2009},
	pages = {104416},
}

@article{chakraborty_highly_2025,
	title = {Highly efficient non-relativistic {Edelstein} effect in nodal p-wave magnets},
	volume = {16},
	copyright = {2025 The Author(s)},
	issn = {2041-1723},
	url = {https://www.nature.com/articles/s41467-025-62516-0},
	doi = {10.1038/s41467-025-62516-0},
	language = {en},
	number = {1},
	urldate = {2025-10-26},
	journal = {Nature Communications},
	author = {Chakraborty, Atasi and Birk Hellenes, Anna and Jaeschke-Ubiergo, Rodrigo and Jungwirth, Tomás and Šmejkal, Libor and Sinova, Jairo},
	month = aug,
	year = {2025},
	note = {Publisher: Nature Publishing Group},
	pages = {7270},
}

@misc{daghofer_altermagnetic_2025,
	title = {Altermagnetic polarons},
	url = {http://arxiv.org/abs/2506.03261},
	doi = {10.48550/arXiv.2506.03261},
	urldate = {2025-10-22},
	publisher = {arXiv},
	author = {Daghofer, Maria and Wohlfeld, Krzysztof and Brink, Jeroen van den},
	month = jun,
	year = {2025},
	note = {arXiv:2506.03261 [cond-mat]},
}

@article{brekke_minimal_2024,
	title = {Minimal {Models} and {Transport} {Properties} of {Unconventional} p -{Wave} {Magnets}},
	volume = {133},
	issn = {0031-9007, 1079-7114},
	url = {https://link.aps.org/doi/10.1103/PhysRevLett.133.236703},
	doi = {10.1103/PhysRevLett.133.236703},
	language = {en},
	number = {23},
	urldate = {2025-10-15},
	journal = {Physical Review Letters},
	author = {Brekke, Bjørnulf and Sukhachov, Pavlo and Giil, Hans Gløckner and Brataas, Arne and Linder, Jacob},
	month = dec,
	year = {2024},
	pages = {236703},
}

@article{bourges_loop_2022,
	title = {Loop currents in quantum matter},
	volume = {22},
	issn = {1878-1535},
	url = {http://arxiv.org/abs/2103.13295},
	doi = {10.5802/crphys.84},
	number = {S5},
	urldate = {2025-09-08},
	journal = {Comptes Rendus. Physique},
	author = {Bourges, Philippe and Bounoua, Dalila and Sidis, Yvan},
	month = may,
	year = {2022},
	note = {arXiv:2103.13295 [cond-mat]},
	pages = {7--31},
}

@article{varma_pseudogap_1999,
	title = {Pseudogap {Phase} and the {Quantum}-{Critical} {Point} in {Copper}-{Oxide} {Metals}},
	volume = {83},
	copyright = {http://link.aps.org/licenses/aps-default-license},
	issn = {0031-9007, 1079-7114},
	url = {https://link.aps.org/doi/10.1103/PhysRevLett.83.3538},
	doi = {10.1103/PhysRevLett.83.3538},
	language = {en},
	number = {17},
	urldate = {2025-09-05},
	journal = {Physical Review Letters},
	author = {Varma, C. M.},
	month = oct,
	year = {1999},
	pages = {3538--3541},
}

@article{fauque_magnetic_2006,
	title = {Magnetic {Order} in the {Pseudogap} {Phase} of {High}-\$\{{T}\}\_\{{C}\}\$ {Superconductors}},
	volume = {96},
	url = {https://link.aps.org/doi/10.1103/PhysRevLett.96.197001},
	doi = {10.1103/PhysRevLett.96.197001},
	number = {19},
	urldate = {2025-05-16},
	journal = {Physical Review Letters},
	author = {Fauqué, B. and Sidis, Y. and Hinkov, V. and Pailhès, S. and Lin, C. T. and Chaud, X. and Bourges, P.},
	month = may,
	year = {2006},
	note = {Publisher: American Physical Society},
	pages = {197001},
}

@article{weber_phase_2014,
	title = {Phase {Diagram} of a {Three}-{Orbital} {Model} for {High}- {Tc} {Cuprate} {Superconductors}},
	volume = {112},
	issn = {0031-9007, 1079-7114},
	url = {https://link.aps.org/doi/10.1103/PhysRevLett.112.117001},
	doi = {10.1103/PhysRevLett.112.117001},
	language = {en},
	number = {11},
	urldate = {2024-01-31},
	journal = {Physical Review Letters},
	author = {Weber, Cédric and Giamarchi, T. and Varma, C. M.},
	month = mar,
	year = {2014},
	pages = {117001},
}

@article{wu_fermi_2007,
	title = {Fermi liquid instabilities in the spin channel},
	volume = {75},
	url = {https://link.aps.org/doi/10.1103/PhysRevB.75.115103},
	doi = {10.1103/PhysRevB.75.115103},
	number = {11},
	urldate = {2025-02-12},
	journal = {Physical Review B},
	author = {Wu, Congjun and Sun, Kai and Fradkin, Eduardo and Zhang, Shou-Cheng},
	month = mar,
	year = {2007},
	note = {Publisher: American Physical Society},
	pages = {115103},
}

@article{pomeranchuk_stability_1958,
	title = {On the {Stability} of a {Fermi} {Liquid}},
	volume = {8},
	language = {en},
	journal = {Sov. Phys. JETP},
	author = {Pomeranchuk, I. I.},
	year = {1958},
	pages = {361},
}

@article{varma_theory_2006,
	title = {Theory of the pseudogap state of the cuprates},
	volume = {73},
	copyright = {http://link.aps.org/licenses/aps-default-license},
	issn = {1098-0121, 1550-235X},
	url = {https://link.aps.org/doi/10.1103/PhysRevB.73.155113},
	doi = {10.1103/PhysRevB.73.155113},
	language = {en},
	number = {15},
	urldate = {2024-08-27},
	journal = {Physical Review B},
	author = {Varma, C. M.},
	month = apr,
	year = {2006},
	pages = {155113},
}

@article{weber_orbital_2009,
	title = {Orbital {Currents} in {Extended} {Hubbard} {Models} of {High}-{\textless}span class="nocase"{\textgreater}{Tc}{\textless}/span{\textgreater} {Cuprate} {Superconductors}},
	volume = {102},
	issn = {0031-9007, 1079-7114},
	url = {https://link.aps.org/doi/10.1103/PhysRevLett.102.017005},
	doi = {10.1103/PhysRevLett.102.017005},
	language = {en},
	number = {1},
	urldate = {2024-01-31},
	journal = {Physical Review Letters},
	author = {Weber, Cédric and Läuchli, Andreas and Mila, Frédéric and Giamarchi, Thierry},
	month = jan,
	year = {2009},
	pages = {017005},
}

@article{leeb_spontaneous_2024,
	title = {Spontaneous {Formation} of {Altermagnetism} from {Orbital} {Ordering}},
	volume = {132},
	issn = {0031-9007, 1079-7114},
	url = {https://link.aps.org/doi/10.1103/PhysRevLett.132.236701},
	doi = {10.1103/PhysRevLett.132.236701},
	language = {en},
	number = {23},
	urldate = {2024-07-04},
	journal = {Physical Review Letters},
	author = {Leeb, Valentin and Mook, Alexander and Šmejkal, Libor and Knolle, Johannes},
	month = jun,
	year = {2024},
	pages = {236701},
}

@article{chakravarty_hidden_2001,
	title = {Hidden order in the cuprates},
	volume = {63},
	issn = {0163-1829, 1095-3795},
	url = {https://link.aps.org/doi/10.1103/PhysRevB.63.094503},
	doi = {10.1103/PhysRevB.63.094503},
	language = {en},
	number = {9},
	urldate = {2024-01-31},
	journal = {Physical Review B},
	author = {Chakravarty, Sudip and Laughlin, R. B. and Morr, Dirk K. and Nayak, Chetan},
	month = jan,
	year = {2001},
	pages = {094503},
}

@article{kung_numerical_2014,
	title = {Numerical exploration of spontaneous broken symmetries in multiorbital {Hubbard} models},
	volume = {90},
	issn = {1098-0121, 1550-235X},
	url = {https://link.aps.org/doi/10.1103/PhysRevB.90.224507},
	doi = {10.1103/PhysRevB.90.224507},
	language = {en},
	number = {22},
	urldate = {2024-01-31},
	journal = {Physical Review B},
	author = {Kung, Y. F. and Chen, C.-C. and Moritz, B. and Johnston, S. and Thomale, R. and Devereaux, T. P.},
	month = dec,
	year = {2014},
	pages = {224507},
}

@article{bulut_instability_2015,
	title = {Instability towards staggered loop currents in the three-orbital model for cuprate superconductors},
	volume = {92},
	issn = {1098-0121, 1550-235X},
	url = {https://link.aps.org/doi/10.1103/PhysRevB.92.195140},
	doi = {10.1103/PhysRevB.92.195140},
	language = {en},
	number = {19},
	urldate = {2024-01-18},
	journal = {Physical Review B},
	author = {Bulut, S. and Kampf, A. P. and Atkinson, W. A.},
	month = nov,
	year = {2015},
	pages = {195140},
}

@article{smejkal_emerging_2022,
	title = {Emerging {Research} {Landscape} of {Altermagnetism}},
	volume = {12},
	issn = {2160-3308},
	url = {https://link.aps.org/doi/10.1103/PhysRevX.12.040501},
	doi = {10.1103/PhysRevX.12.040501},
	language = {en},
	number = {4},
	urldate = {2023-09-12},
	journal = {Physical Review X},
	author = {Šmejkal, Libor and Sinova, Jairo and Jungwirth, Tomas},
	month = dec,
	year = {2022},
	pages = {040501},
}

@article{gonzalez-hernandez_efficient_2021,
	title = {Efficient {Electrical} {Spin} {Splitter} {Based} on {Nonrelativistic} {Collinear} {Antiferromagnetism}},
	volume = {126},
	issn = {0031-9007, 1079-7114},
	url = {https://link.aps.org/doi/10.1103/PhysRevLett.126.127701},
	doi = {10.1103/PhysRevLett.126.127701},
	language = {en},
	number = {12},
	urldate = {2023-09-12},
	journal = {Physical Review Letters},
	author = {González-Hernández, Rafael and Šmejkal, Libor and Výborný, Karel and Yahagi, Yuta and Sinova, Jairo and Jungwirth, Tomáš and Železný, Jakub},
	month = mar,
	year = {2021},
	pages = {127701},
}

@misc{supplement,
    note = {
        The Supplemental Material includes the Bloch Hamiltonians for the 4 site unit cell of the square lattice, the Lieb lattice and the sign conventions for the currents.
        URL: [url inserted by publisher].
    }
}

@article{corticelli2022spin,
  title={Spin-space groups and magnon band topology},
  author={Corticelli, Alberto and Moessner, Roderich and McClarty, Paul A},
  journal={Physical Review B},
  volume={105},
  number={6},
  pages={064430},
  year={2022},
  publisher={APS}
}

@article{aoyama2024piezomagnetic,
  title={Piezomagnetic properties in altermagnetic MnTe},
  author={Aoyama, Takuya and Ohgushi, Kenya},
  journal={Physical Review Materials},
  volume={8},
  number={4},
  pages={L041402},
  year={2024},
  publisher={APS}
}

@article{yu2025odd,
  title={Odd-parity magnetism driven by antiferromagnetic exchange},
  author={Yu, Yue and Lyngby, Magnus B and Shishidou, Tatsuya and Roig, Merc{\`e} and Kreisel, Andreas and Weinert, Michael and Andersen, Brian M and Agterberg, Daniel F},
  journal={Physical Review Letters},
  volume={135},
  number={4},
  pages={046701},
  year={2025},
  publisher={APS}
}

@article{yamada2025metallic,
  title={A metallic p-wave magnet with commensurate spin helix},
  author={Yamada, Rinsuke and Birch, Max T and Baral, Priya R and Okumura, Shun and Nakano, Ryota and Gao, Shang and Ezawa, Motohiko and Nomoto, Takuya and Masell, Jan and Ishihara, Yuki and others},
  journal={Nature},
  volume={646},
  number={8086},
  pages={837--842},
  year={2025},
  publisher={Nature Publishing Group UK London}
}

@article{zhou2025anisotropic,
  title={Anisotropic resistivity of a $ p $-wave magnet candidate CeNiAsO},
  author={Zhou, Honglin and Wang, Muyu and Ma, Xiaoyan and Li, Gang and Shao, Ding-Fu and Liu, Bo and Li, Shiliang},
  journal={arXiv preprint arXiv:2509.07351},
  year={2025}
}

@article{song2025electrical,
  title={Electrical switching of ap-wave magnet},
  author={Song, Qian and Stavri{\'c}, Srdjan and Barone, Paolo and Droghetti, Andrea and Antonenko, Daniil S and Venderbos, J{\"o}rn WF and Occhialini, Connor A and Ilyas, Batyr and Erge{\c{c}}en, Emre and Gedik, Nuh and others},
  journal={Nature},
  pages={1--7},
  year={2025},
  publisher={Nature Publishing Group UK London}
}

@article{affleck_largen_1988,
  title = {Large-n Limit of the {{Heisenberg-Hubbard}} Model: {{Implications}} for High-\$\textbraceleft{{T}}\textbraceright\_\textbraceleft c\textbraceright\$ Superconductors},
  author = {Affleck, Ian and Marston, J. Brad},
  year = 1988,
  journal = {Phys. Rev. B},
  volume = {37},
  number = {7},
  pages = {3774--3777},
  doi = {10.1103/PhysRevB.37.3774}
}

@article{dzyaloshinskii_magnetoelectrical_1960,
  title = {On the Magneto-Electrical Effects in Antiferromagnets},
  author = {Dzyaloshinski{\v i}, I.},
  year = 1960,
  journal = {Soviet physics, JETP}
}

@article{feng_lowenergy_2021,
  title = {Low-Energy Effective Theory and Symmetry Classification of Flux Phases on the Kagome Lattice},
  author = {Feng, Xilin and Zhang, Yi and Jiang, Kun and Hu, Jiangping},
  year = 2021,
  journal = {Phys. Rev. B},
  volume = {104},
  number = {16},
  pages = {165136},
  doi = {10.1103/PhysRevB.104.165136}
}

@article{graham_depthdependent_2024,
  title = {Depth-Dependent Study of Time-Reversal Symmetry-Breaking in the Kagome Superconductor {{AV3Sb5}}},
  author = {Graham, J. N. and Mielke III, C. and Das, D. and Morresi, T. and Sazgari, V. and Suter, A. and Prokscha, T. and Deng, H. and Khasanov, R. and Wilson, S. D. and Salinas, A. C. and Martins, M. M. and Zhong, Y. and Okazaki, K. and Wang, Z. and Hasan, M. Z. and Fischer, M. H. and Neupert, T. and Yin, J.-X. and Sanna, S. and Luetkens, H. and Salman, Z. and Bonf{\`a}, P. and Guguchia, Z.},
  year = 2024,
  journal = {Nat Commun},
  volume = {15},
  number = {1},
  pages = {8978},
  doi = {10.1038/s41467-024-52688-6}
}

@misc{huang_lightinduced_2025,
  title = {Light-Induced {{Odd-parity Magnetism}} in {{Conventional Collinear Antiferromagnets}}},
  author = {Huang, Shengpu and Qin, Zheng and Zhan, Fangyang and Xu, Dong-Hui and Ma, Da-Shuai and Wang, Rui},
  year = 2025,
  number = {arXiv:2507.20705},
  eprint = {2507.20705},
  doi = {10.48550/arXiv.2507.20705},
  archiveprefix = {arXiv}
}

@misc{li_floquet_2025,
  title = {Floquet {{Spin Splitting}} and {{Spin Generation}} in {{Antiferromagnets}}},
  author = {Li, Bo and Shao, Ding-Fu and Kovalev, Alexey A.},
  year = 2025,
  number = {arXiv:2507.22884},
  eprint = {2507.22884},
  doi = {10.48550/arXiv.2507.22884},
  archiveprefix = {arXiv}
}

@misc{lin_oddparity_2025,
  title = {Odd-Parity Altermagnetism through Sublattice Currents: {{From Haldane-Hubbard}} Model to General Bipartite Lattices},
  author = {Lin, Yu-Ping},
  year = 2025,
  number = {arXiv:2503.09602},
  eprint = {2503.09602},
  doi = {10.48550/arXiv.2503.09602},
  archiveprefix = {arXiv}
}

@misc{liu_lightinduced_2025,
  title = {Light-Induced Odd-Parity Altermagnets on Dimerized Lattices},
  author = {Liu, Dongling and Zhuang, Zheng-Yang and Zhu, Di and Wu, Zhigang and Yan, Zhongbo},
  year = 2025,
  number = {arXiv:2508.18360},
  eprint = {2508.18360},
  doi = {10.48550/arXiv.2508.18360},
  archiveprefix = {arXiv}
}

@misc{luo_spin_2026,
  title = {Spin {{Group Symmetry Criteria}} for {{Odd-parity Magnets}}},
  author = {Luo, Xun-Jiang and Hu, Jin-Xin and Hu, Meng-Li and Law, K. T.},
  year = 2026,
  number = {arXiv:2510.05512},
  eprint = {2510.05512},
  doi = {10.48550/arXiv.2510.05512},
  archiveprefix = {arXiv}
}

@article{marston_largen_1989,
  title = {Large-n Limit of the {{Hubbard-Heisenberg}} Model},
  author = {Marston, J. Brad and Affleck, Ian},
  year = 1989,
  journal = {Phys. Rev. B},
  volume = {39},
  number = {16},
  pages = {11538--11558},
  doi = {10.1103/PhysRevB.39.11538}
}

@misc{meier_antialtermagnetism_2025,
  title = {({{Anti-}}){{Altermagnetism}} from {{Orbital Ordering}} in the {{Ruddlesden-Popper Chromates Sr}}\$\_\textbraceleft n+1\textbraceright\${{Cr}}\$\_n\${{O}}\$\_\textbraceleft 3n+1\textbraceright\$},
  author = {Meier, Quintin N. and Carta, Alberto and Ederer, Claude and Cano, Andres},
  year = 2025,
  number = {arXiv:2502.01515},
  eprint = {2502.01515},
  doi = {10.48550/arXiv.2502.01515},
  archiveprefix = {arXiv}
}

@article{mielke_timereversal_2022,
  title = {Time-Reversal Symmetry-Breaking Charge Order in a Kagome Superconductor},
  author = {Mielke, C. and Das, D. and Yin, J.-X. and Liu, H. and Gupta, R. and Jiang, Y.-X. and Medarde, M. and Wu, X. and Lei, H. C. and Chang, J. and Dai, Pengcheng and Si, Q. and Miao, H. and Thomale, R. and Neupert, T. and Shi, Y. and Khasanov, R. and Hasan, M. Z. and Luetkens, H. and Guguchia, Z.},
  year = 2022,
  journal = {Nature},
  volume = {602},
  number = {7896},
  pages = {245--250},
  doi = {10.1038/s41586-021-04327-z}
}

@article{sato_altermagnetic_2024,
  title = {Altermagnetic {{Anomalous Hall Effect Emerging}} from {{Electronic Correlations}}},
  author = {Sato, Toshihiro and Haddad, Sonia and Fulga, Ion Cosma and Assaad, Fakher F. and {van den Brink}, Jeroen},
  year = 2024,
  journal = {Phys. Rev. Lett.},
  volume = {133},
  number = {8},
  pages = {086503},
  doi = {10.1103/PhysRevLett.133.086503}
}

@article{smejkal_conventional_2022,
  title = {Beyond {{Conventional Ferromagnetism}} and {{Antiferromagnetism}}: {{A Phase}} with {{Nonrelativistic Spin}} and {{Crystal Rotation Symmetry}}},
  author = {{\v S}mejkal, Libor and Sinova, Jairo and Jungwirth, Tomas},
  year = 2022,
  journal = {Phys. Rev. X},
  volume = {12},
  number = {3},
  pages = {031042},
  doi = {10.1103/PhysRevX.12.031042}
}

@article{tazai_chargeloop_2023,
  title = {Charge-Loop Current Order and {{Z3}} Nematicity Mediated by Bond Order Fluctuations in Kagome Metals},
  author = {Tazai, Rina and Yamakawa, Youichi and Kontani, Hiroshi},
  year = 2023,
  journal = {Nat Commun},
  volume = {14},
  number = {1},
  pages = {7845},
  doi = {10.1038/s41467-023-42952-6}
}

@article{varma_nonfermiliquid_1997,
  title = {Non-{{Fermi-liquid}} States and Pairing Instability of a General Model of Copper Oxide Metals},
  author = {Varma, C. M.},
  year = 1997,
  journal = {Phys. Rev. B},
  volume = {55},
  number = {21},
  pages = {14554--14580},
  doi = {10.1103/PhysRevB.55.14554}
}

@misc{zeng_oddparity_2025,
  title = {The Odd-Parity Altermagnetism: {{A}} Spin Group Study},
  author = {Zeng, Minghuan and Qin, Zheng and Qin, Ling and Feng, Shiping and Xu, Dong-Hui and Wang, Rui},
  year = 2025,
  number = {arXiv:2507.09906},
  eprint = {2507.09906},
  doi = {10.48550/arXiv.2507.09906},
  archiveprefix = {arXiv}
}

@misc{zhu_floquet_2025,
  title = {Floquet Odd-Parity Collinear Magnets},
  author = {Zhu, Tongshuai and Zhou, Di and Wang, Huaiqiang and Ruan, Jiawei},
  year = 2025,
  number = {arXiv:2508.02542},
  eprint = {2508.02542},
  doi = {10.48550/arXiv.2508.02542},
  archiveprefix = {arXiv}
}

@misc{zhuang_oddparity_2025,
  title = {Odd-{{Parity Altermagnetism Originated}} from {{Orbital Orders}}},
  author = {Zhuang, Zheng-Yang and Zhu, Di and Liu, Dongling and Wu, Zhigang and Yan, Zhongbo},
  year = 2025,
  number = {arXiv:2508.18361},
  eprint = {2508.18361},
  doi = {10.48550/arXiv.2508.18361},
  archiveprefix = {arXiv}
}
%extra ref are paper specific references which are papers: SM,..
%bib is automatic file of zotero
%zotero is an experted file of zotero for finalizing

%\newpage
\clearpage

\onecolumngrid
\begin{center}
\textbf{\large End Matter}
\end{center}
\twocolumngrid
\appendix

\sectionPRL{\thesection.~Minimal 1D model}
\label{A:sec:minimal 1D}
The Hamiltonian
\begin{align}
    H_{1D} =& \sum_{j,s=\pm} sm (c_{j,B}^\dag c_{j,B} -c_{j,A}^\dag c_{j,A}) - t c_{j,B}^\dag c_{j,A} 
    \nonumber \\
    &- t \left(\e^{-\ii \Phi/2}c_{j+1,B}^\dag c_{j,B} - \e^{\ii \Phi/2} c_{j+1,A}^\dag c_{j,A}\right)
\end{align}
has three inequivalent, spin-independent hoppings. In momentum space the Bloch Hamiltonian $h_s(k) = \vc{\sigma} \cdot \vc{d}^s$ is block diagonal in spin and is conveniently written in terms of Pauli matrices $\vc{\sigma} = (\sigma_0,\sigma_x,\sigma_y,\sigma_z)$, where 
\begin{align}
    \vc{d}^s =
    \begin{pmatrix}
        -2t \cos k \cos \frac{\Phi}{2} \\
        -t \\
        0 \\
        -2t\sin k \sin \frac{\Phi}{2}-sm
    \end{pmatrix}
\end{align}
Therefore we can calculate the eigenvectors and eigenvalues $\epsilon^\pm_s(k) = d_0 \pm \sqrt{d_x^2 + {d_z^s}^2}$ exactly.
%and the eigenstates
% \begin{equation}
%     |u^\pm_s(k) \rangle = \frac{1}{\sqrt{\mathcal{N}^\pm_s}} \left(d^s_z \pm \sqrt{d_x^2 + d_z^{s2}}, d_x\right)
% \end{equation}
% where $\mathcal{N}^\pm_s = d_x^2 + \left(d_z^s \pm \sqrt{d_x^2 + d_z^{s2}}\right)^2$, exactly. 
% The Edelstein susceptibility vanishes, because in 1D there are no crossings of spin-up and down bands except at the band edges. 

\begin{figure}
    \centering
    \includegraphics[width=\columnwidth]{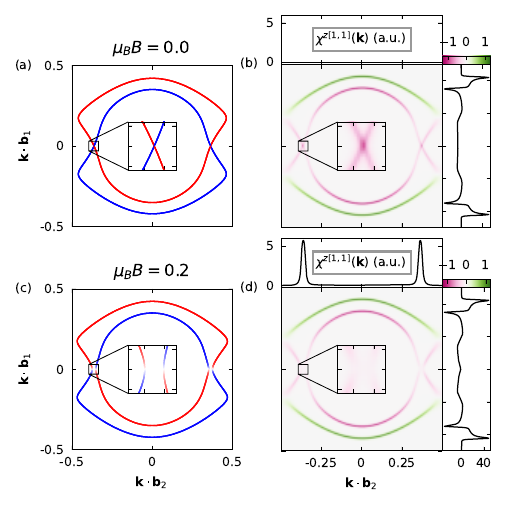}
    \caption{Origin of the magnetic Edelstein effect. The Fermi surface colored by spin (a,c) and the momentum-resolved Edelstein susceptibility (b,d) of a collinear $p$-wave magnet for zero magnetic field (a,b) and finite magnetic field (b,d). A magnetic field lifts the band crossings in the band structure. The sidepanels are histrogram plots of the momentum-resolved Edelstein susceptibility, i.e., $\sum_{k_x=\pm k_y} \chi^{z,[1,1]}(\vc{k})$. The top panel of (b) shows that the positive contributions (green) and negative contributions (purple) of the momentum-resolved Edelstein susceptibility cancel for each momentum exactly. At finite  magnetic field (d) the avoided crossing does not lead to a contribution to the momentum-resolved Edelstein susceptibility. Hence, the contribution of the outer bands (green) is uncompensated at this momentum which leads to Gaussian peaks in the top panel of (d).}
    \label{fig:MEE}
\end{figure}

\sectionPRL{\thesection.~Magnetic Edelstein effect}
\label{A:sec:magnetic Edelstein effect}
\label{A:sec:Edelstein effect 0}
\label{A:sec:Zeeman field}
% \subsection{Vanishing Edelstein effect in spin-commuting Hamiltonians}
% \label{A:sec:Edelstein effect 0}
In the case of spin commuting Hamiltonians $[\sigma^i,H] = 0$, the spin texture $\langle u_n(\vc{k}) | \sigma^i|u_m(\vc{k})\rangle = s^i_n \delta_{n,m}$ is diagonal. This reduces the band summation in \eqref{eq:Edelstein suscetibility} to a single band, which simplifies the terms
\begin{align}
    \langle u_n(\vc{k}) | \partial_j h(\vc{k}) |u_n(\vc{k})\rangle &= \partial_j \epsilon_n(\vc{k}) 
    \\
    \Re \left[\left|G_n(\vc{k})\right|^2 - G_n(\vc{k})^2 \right] &= \frac{\pi}{\Gamma} \delta\left(\epsilon_n(\vc{k})\right).
\end{align}
Hence, the Edelstein susceptibility can be written as a pure Fermi surface term 
\begin{align}
\chi^{ij} \Gamma =& \frac{e \pi}{V} \sum_{\vc{k},n} s^i_n \partial_j \epsilon_n(\vc{k}) \delta(\epsilon_n(\vc{k}))
\\
=& \frac{e \pi}{(2\pi)^d} \sum_n s^i_n \oint_{\epsilon_n(\vc{k})=0} \d k \frac{\partial_j \epsilon_n(\vc{k})}{|\nabla \epsilon_n(\vc{k})|}
, 
\label{eq:Edelstein suscetibility FS}
\end{align}
where $d$ is the dimension. Note, that this is identical to the intraband part of the susceptibility, which is for spin-commuting Hamiltonians the only contributing part \cite{garate_nonadiabatic_2009,chakraborty_highly_2025}.

\begin{figure}[t]
    \centering
    \includegraphics[width=\columnwidth]{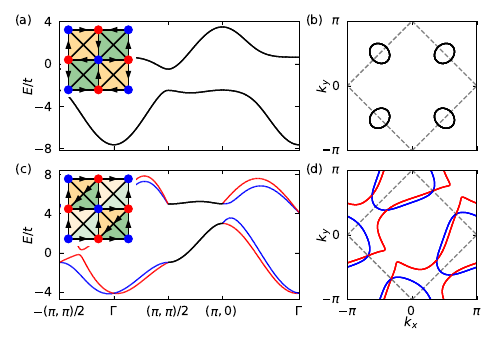}
    \caption{Additional examples of LCOs coexisting with Neél AFM states. (a,c) shows the band structure along a high symmetry line where the inset depicts the state and (b,d) a representative Fermi surface of the state. (a,b) is based on the historic LCO introduced as $d$-density wave by Ref.~\cite{chakravarty_hidden_2001}. It is a conventional AFM even though a LCO is present such that $[C_2||\vc{t}]$ is broken, because $[\T||\T \P]$ is preserved. The state shown in (c,d) constitutes a second example for a hidden orbital ferrimagnet.}
    \label{fig:LCOs_square}
\end{figure}

The closed line integral of \eqref{eq:Edelstein suscetibility FS} vanishes for any Fermi surface topography. Note that even open Fermi pockets can be considered closed line integrals on the Brillouin zone manifold. The integrand is identical to the $i$-th component of the normal unit vector $n_j = \hat{\vc{e}}_j \cdot \vc{n} = \partial_j \epsilon_n(\vc{k})/|\nabla \epsilon_n(\vc{k})|$. Hence, the integral vanishes due to Gauss's law
\begin{align}
    &\oint_{\epsilon_n(\vc{k})=0} \d k \frac{\partial_j \epsilon_n(\vc{k})}{|\nabla \epsilon_n(\vc{k})|}
    =
    \oint_{\epsilon_n(\vc{k})=0} \d k \hat{\vc{e}}_j \cdot \vc{n}
    \nonumber
    \\
    =&
    \int_{\epsilon_n(\vc{k})\leq 0} \d^d k \nabla \cdot \hat{\vc{e}}_j
    =
    0.
\end{align}
This proofs rigorously that a modulated spin polarization is needed to observe the Edelstein effect, which has been mentioned but not explained before in Ref.~\cite{chakraborty_highly_2025}. For a spin-degeneracy preserving Fermi surface each sheet of Fermi surfaces vanishes individually, see Fig.~\ref{fig:MEE}~(b).

\begin{figure}
    \centering
    \includegraphics[width=\columnwidth]{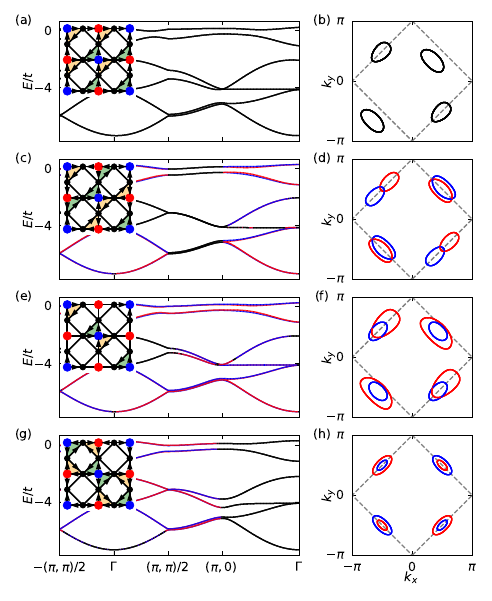}
    \caption{Unconventional magnetic phases on the Lieb lattice. (a,b) is an inversion-broken AFM. The LCO is identical with the intra unit cell time reversal violating state $\Theta_{II}$, introduced by Varma in 2006 \cite{varma_theory_2006}. The staggered version of this LCO, i.e., with an ordering vector $\vc{Q}=(\pi,\pi)$, leads to a $p$-wave magnet, shown in (c,d). A superposition of the orderings of the former two LCOs, i.e., the LCO forms exclusively around the spin-up sublattice, leads the a hidden orbital ferrimagnet (e,f). On the Lieb lattice already a $\vc{Q}=(\pi,\pi)$ LCO can induce an AM, see (g,h). }
    \label{fig:LCOs_lieb}
\end{figure}

% \subsection{Effect of a Zeeman field}
% \label{A:sec:Zeeman field}
A small magnetic field $\vc{B}$ perpendicular to the Néel vector generates a non-zero Edelstein effect. The relevant coupling is the Zeeman term $\mu_B \vc{\sigma} \cdot \vc{B}$ ($\mu_B$ is the Bohr magneton). To understand the reason, it is instructive to study the momentum-resolved Edelstein susceptibility
\begin{align}
    \chi^{ij}(\vc{k}) = \frac{e}{2\pi V} & \Re \sum_{n,m}  \left[G_n(\vc{k})G_m^*(\vc{k}) - G_n(\vc{k}) G_m(\vc{k}) \right]
    \nonumber \\
    &\times 
    \langle u_n(\vc{k}) | \sigma^i|u_m(\vc{k})\rangle \langle u_m(\vc{k}) | \partial_j h(\vc{k}) |u_n(\vc{k})\rangle
\end{align}
which satisfies $\chi^{ij} = \chi^{ij}(\vc{k})$ and is shown in Fig.~\ref{fig:MEE}.

Generically, the Fermi surface of a collinear $p$-wave magnets consists of at least 2 Fermi sheets, one of each spin character, see Fig.~\ref{fig:MEE}~(a). A Zeeman field now leads to a small hybridization gap between these bands [Fig.~\ref{fig:MEE}~(b)]. However, away from the hybridization region, the Fermi surface remains unaffected by the magnetic field, such that at these momenta the momentum-resolved Edelstein susceptibility does not contribute to a net Edelstein effect, see the top sidepanels of Fig.~\ref{fig:MEE}~(b,d). In the hybridization region the momentum-resolved Edelstein susceptibility vanishes, because of the hybridization [Fig.~\ref{fig:MEE}~(d)]. Hence, the momentum-resolved Edelstein susceptibility of the parts of the Fermi surface at shifted momenta is uncompensated and can lead to a weak magnetic moment perpendicular to the applied magnetic field.

\sectionPRL{\thesection.~Square lattice}
The famous $d$-wave $(\pi,\pi)$-LCO of Ref.~\cite{chakravarty_hidden_2001}, shown in Fig.~\ref{fig:LCOs_square}~(a), leads to a conventional AFM, because of the presence of inversion symmetry and time reversal times translation. It is described by \eqref{eq:H_square} with
\begin{align}
    \vc{d}_{d^2} (\vc{k}) = 2 z_{d^2} \left[\cos k_x - \cos k_y \right] \hat{\vc{e}}_y.
\end{align}
The LCO shown in Fig.~\ref{fig:LCOs_square}~(c) leads to a hidden orbital ferrimagnet and is a superposition of $z_{p^2_{[1,1]}}$ and $-z_{p^1_{[1,1]}}$. 

\sectionPRL{\thesection.~Lieb lattice}
\label{A:sec:lieb lattice}
The three-band Emery model is a minimal model for the cuprates \cite{emery_theory_1987}. The Hamiltonian
\begin{align}
\mathcal{H}=&\sum_{i\alpha s}\varepsilon_{\alpha}n_{i\alpha s}+\sum_{i\alpha j\beta s}t_{i\alpha j\beta}\mathbf{c}_{i\alpha s}^\dagger \mathbf{c}_{j\beta s} 
\label{eq:hamiltonian Emery}
\end{align}
consists of a Lieb lattice with $d_{x^2-y^2}$ Cu orbitals ($\alpha,\beta = d$) as central sites which are surrounded by O $p_x$ and $p_y$ orbitals ($\alpha,\beta = p_x, p_y$), see Fig.~\ref{fig:LCOs_lieb} or Ref.~\cite{li_exploring_2025}. We consider Cu-O hoppings $t_{pd}$ and O-O hoppings $t_{pp} = t_{pd}/2$. We set the orbital onsite energy to $\varepsilon_d=0$ and $\varepsilon_p = -3 t_{pd}$, mimicking the band structure of the cuprates. Additionally, we consider the commonly observed $(\pi,\pi)$-AFM order on the Cu atoms quantified by $m$ and the $(0,0)$ and $(\pi,\pi)$-LCOs shown in Fig.~\ref{fig:LCOs_lieb}. The full Bloch Hamiltonian is provided in the SM~\cite{supplement}.

\end{document}